\documentstyle[12pt]{article}                      

\newcommand{\beq}{\begin{equation}}
\newcommand{\eeq}{\end{equation}}
\newcommand{\bq}{\begin{quotation}}
\newcommand{\eq}{\end{quotation}}
\newcommand{\bc}{\begin{center}}
\newcommand{\ec}{\end{center}}

\begin{document} 

\title{{\sc  Limits of time in cosmology}}
\author{
{\sc S.E. Rugh\footnote{Symposion,``The Socrates Spirit", Section for Philosophy and the Foundations of Physics, Helleb\ae kgade 27, Copenhagen N, Denmark
({\em e-mail: rugh@symposion.dk).}}  \addtocounter{footnote}{5} and H. Zinkernagel\footnote{Department of Philosophy I, University of Granada, 18071
Granada, Spain ({\em e-mail: zink@ugr.es}).}  }}
\date{}

\maketitle

\begin{quotation}

\noindent
To appear in {\em The Philosophy of Cosmology}; edited by K. Chamcham, J. Silk, J. Barrow and S. Saunders. Cambridge University Press, 2016.

\vspace{1cm}

\noindent
\small
{\bf Abstract}  We provide a discussion of some main ideas in our project about the physical foundation of the time concept in cosmology. It is standard to point to the Planck scale (located
at $\sim 10^{-43}$ seconds after a fictitious  ``Big Bang" point) as a limit for how far back we may 
extrapolate the standard cosmological model. In our work we have suggested that there
are several other (physically motivated) interesting limits -- located at least thirty orders of magnitude before the Planck time  --  where the physical basis of the
cosmological model and its time concept is progressively weakened. Some of these limits are connected to phase transitions in the early universe which gradually undermine the
notion of 'standard clocks' widely employed in cosmology.
Such considerations lead to a {\em {scale problem}} for time which becomes particularly acute above the electroweak phase transition (before $\sim 10^{-11}$ seconds). Other limits are due to problems of building up a cosmological reference frame, or even contemplating a sensible notion of proper time, if the early universe constituents become too quantum. This {\em {quantum problem}} for time arises e.g. if a pure quantum phase is contemplated at the beginning of inflation at, say, $\sim 10^{-34}$ seconds.

\end{quotation} 

\normalsize

\newpage

\section{Introduction}

\label{Introduction}

What does time mean in cosmology? Are there any physical conditions which must be satisfied in order to speak about cosmic time? If so, how far back can time be extrapolated while still maintaining it as a well-defined physical concept? We have studied these questions in a series of papers
over the last ten years. The present manuscript is a summary of some main points from our investigations, as well as some further considerations regarding time in cosmology.

It is standard to assume that a number of important events took place in
the first tiny fractions of a second `after' the big bang. For instance,
the universe is thought to have been in a quark-gluon phase between
$\sim 10^{-11} - 10^{-5}$ seconds, whereas the fundamental material
constituents are massless due to the electroweak (Higgs) transition at
times earlier than $\sim 10^{-11}$ seconds. A phase of inflation is
envisaged (in some models) to have taken place around $\sim 10^{-34}$ seconds
after the big bang. A rough summary of the phases of the early universe
is given in the figure:\footnote{For a detailed discussion of (the assumptions behind) this figure and the epochs indicated, see also Rugh and Zinkernagel  (2009).}

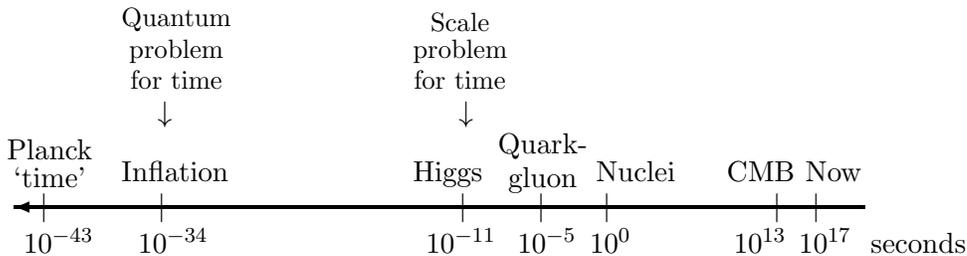
\begin{figure}[h]
\small
\label{timeline}

\setlength{\unitlength}{.87cm}                                                
\begin{picture}(4.5,4.5)(-1.5,-.3)

\thicklines

\put(12.8,1){\vector(-1,0){13}}

\put(.2,.88){{\bf $|$}}
\put(-.3,1.3){{\shortstack {Planck \\ `time'}}}
\put(-.05,.3){{$10^{-43}$}}

\put(2,.88){{\bf $|$}}
\put(1.45,1.4){{Inflation}}
\put(1.7,.3){{$10^{-34}$}}

\put(2,2.3){{\bf $\downarrow$}}
\put(1.45,2,8){\footnotesize{\shortstack{Quantum \\ problem \\ for time}}}

\put(6.6,.88){{\bf $|$}}
\put(5.9,1.4){{Higgs}}
\put(6.1,.3){{$10^{-11}$}}

\put(6.6,2.3){{\bf $\downarrow$}}
\put(5.9,2.8){\footnotesize{\shortstack{Scale \\ problem \\ for time}}}

\put(7.8,.88){{\bf $|$}}
\put(7.2,1.3){{\shortstack{Quark- \\ gluon}}}
\put(7.5,.3){{$10^{-5}$}}

\put(8.8,.88){{\bf $|$}}
\put(8.7,1.4){{Nuclei}}
\put(8.6,.3){{$10^{0}$}}

\put(11.4,.88){{\bf $|$}}
\put(10.7,1.4){{CMB}}
\put(10.8,.3){{$10^{13}$}}

\put(12,.88){{\bf $|$}}
\put(11.9,1.4){{Now}}
\put(11.8,.3){{$10^{17}$}}

\put(12.9,.3){{seconds}}

\end{picture}
\caption{{\small Contemplated phases of the early universe. 
The indicated quantum and scale problems for time are discussed in the text. } }
\end{figure}

What could be wrong (or at least problematic) with this backward extrapolation from now?
A main point is that {\em {physical}} time in relativity theory, in contrast to a purely mathematical parameter with the label $t$, is bound up with the notion of proper time.\footnote{Proper time along a (timelike or lightlike) world line (the path of a particle in 4-dimensional spacetime) can be
thought of as the time measured by a  ``standard" clock along that world line.}
For example, Misner, Thorne and Wheeler (1973, p. 813) write:
\begin{quote} 
...proper time is the most physically significant, most
physical real time we know. It corresponds to the ticking of physical
clocks and measures the natural rhythms of actual events.
\end{quote}
The connection between physical time and proper time leads to two kinds of problems
 for the backward extrapolation.
The first of these follows from the fact that proper time is closely related to physical clocks or processes. The nature (and availability) 
of such clocks or processes changes as we go back in time. The problem in this regard was hinted at by Misner, Thorne and Wheeler in connection with a discussion of whether a singularity occurs at a finite past proper time. They note that no actual clock can 
adequately time the earliest moments of the universe:

\begin{quote} 
Each actual clock has its ``ticks" discounted by a suitable factor -
$3*10^{7}$ seconds per orbit from the Earth-sun system, $1.1*10^{-10}$
seconds per oscillation for the Cesium transition, etc. Since no single
clock (because of its finite size and strength) is conceivable all the
way back to the singularity, a statement about the {\em{proper time}} since the
singularity involves the concept of an infinite sequence of successively
smaller and sturdier clocks with their ticks then discounted and added.
[...] ...finiteness [of the age of the universe] would be judged by
counting the number of discrete ticks on {\em realizable clocks}, not by
accessing the weight of unrealizable mathematical abstractions. [Our emphasis]
\vspace{-.2cm}
\begin{flushright}
Misner, Thorne and Wheeler (1973, p. 814)
\end{flushright}
\end{quote}

The authors' discussion regarding this quote seems to imply that the progressively more extreme physical conditions, as we extrapolate the standard cosmological model backwards, demand a succession of gradually more fine-grained clocks to give meaning to (or provide a physical basis of) the time of each of the 
epochs.\footnote{Regarding, Misner, Thorne and Wheeler's examples in the quote, it is clear that
one has to distinguish between 
{\em how fine-grained} a  clock is (its precision) and {\em when} (in which cosmological
epoch) such a clock could in principle be realized. For instance, no stable Cesium atoms -- let alone real functioning Cesium clocks -- can exist before the time of decoupling of radiation and matter, about 380,000 years after the big bang.}
 In this spirit, our view is
that a minimal requirement for having a physical notion of time (with a scale) is that it must be possible to find physical processes (what we call `cores of clocks') with a sufficiently fine-grained duration in the physics envisaged in the various epochs of cosmic history. As we shall discuss below, this requirement of linking time to conceivable cores of clocks leads to a {\em {scale problem}} for time, since it becomes progressively more difficult to identify physical processes with a well-defined
(and non-zero) duration in the very early universe.

A second kind of  problem with the backward extrapolation 
follows since proper time is defined in terms of (possible) particle world lines or trajectories. Within the standard cosmological model, there is a privileged set of such world lines since matter on large scales is assumed to move in a highly ordered manner (allowing for the identification of a 
comoving reference frame  and a global cosmic time equal to
 the proper time of any comoving observer). As we shall discuss, this implies that the notion of cosmic time is closely related to the so-called Weyl principle. Problems with the notion of  a global cosmic time 
 may arise if a privileged  set of world lines becomes 
difficult  to identify, e.g. in the very early universe
above the electroweak (Higgs) phase transition or in a (complicated)
inhomogeneous universe.

A more serious problem for time 
(which is a problem 
even for a local definition of time) arises if a point is reached in the backward extrapolation where the  world lines themselves can no longer be identified. 
In particular, this appears to be the case if some point is contemplated, e.g. at the onset of inflation, where all constituents of the universe are of a quantum nature, leading to what can be called the {\em {quantum problem}} of time. Note that this problem arises roughly ten orders of magnitude 
 `before' (in the backward extrapolation from now) reaching a possible quantum gravity epoch, and so before hitting the usual problem of time in quantum gravity models.

In the following, we first outline the scale problem for time and the close relation between time and clocks. We then address the relation between time and world lines in the set-up of the standard cosmological model. We briefly indicate how this relation may lead to problems for a global (cosmic) time concept in the very early
universe above the electroweak phase transition or in a (complicated)
 inhomogeneous universe. We finally discuss the more serious local (quantum) problem for time 
in relation to the problem of identifying  individual world lines.

\section{Time and clocks}

\label{TimeAndClocks}

The idea that time is dependent on change and/or motion is called relationism. It has been defended by classic thinkers like Aristotle and Leibniz, and in modern times by physicists like Barbour, Smolin and Rovelli. In our version of relationism, we argue in favour of a `time-clock' relation which asserts that time, in order to have a physical basis, must be understood in relation to physical processes which act as `cores' of clocks (Rugh and Zinkernagel 2005, 2009, see also Zinkernagel 2008). In the cosmological context, the time-clock relation implies that a necessary physical condition for {\em interpreting}  the $t$ parameter of the standard Friedmann-Lema\^{\i}tre-Robertson-Walker (FLRW) model as cosmic time in some `epoch' of the universe is the (at least possible) existence of a physical process which can function as a core of a clock in the `epoch' in question. In particular, we have suggested that in order to make the interpretation
$$ t \leftrightarrow \mbox{time}, $$
at a specific cosmological
`epoch', the physical process acting as the core of a clock should 1)
have a well-defined duration which is sufficiently
fine-grained to `time' the epoch in question; and 2) be a process which
could conceivably take place among the material constituents available
in the universe at this epoch.

The time-clock relation is in conformity with how time is employed in
cosmology although cosmologists often formulate themselves in
operationalist terms -- that is, invoking observers measuring on factual
clocks. For
instance, Peacock (1999, p. 67) writes concerning the FLRW model:

\begin{quote}
We can define a global time coordinate $t$, which is the time measured
by clocks of these observers -- i.e. $t$ is the proper time measured by
an observer at rest with respect to the local matter distribution. 
\end{quote}

\noindent
While this reference to clocks (or `standard' clocks) carried by 
comoving observers is widely made in cosmology textbooks, there is
usually no discussion concerning the origin and nature of these clocks.
Part of the motivation for our investigations has been to provide a
discussion of this kind.

The standard definition of the global time coordinate to which Peacock refers -- and, in general, the question of how to make the $t$-time identification -- can be read in at least two different ways: 1) Actual clocks should be available (operationalism); or 2) rudiments (cores)
of clocks with a well-defined duration 
should, in principle, be present (time-clock relation). Clearly, the first possibility is not an option in the very early universe where no actual clocks, let alone observers and measurements, are available. As we shall see below, the viability of the second option depends upon the availability of 
physical processes with well-defined (and non-zero) duration.

We attempt to develop a position on the time concept which represents a departure
from operationalism in several ways: (i) Time cannot be defined (reductively) in terms of clocks (since clocks and measurements depend on the time concept); (ii) no actual clocks are needed, we allow reference to possible (counterfactual) clocks, compatible with the physics of the epoch in question; (iii) we
attempt to construct the {\em cores} of clocks out of available physics,
but do not require that this core should be associated with a counter
mechanism that could transform it into a real functioning clock; and (iv) we
do not require the existence of observers and actual measurements. 
Nevertheless, the above formulated criterion for the $t \leftrightarrow$ time interpretation of being able to identify a process with a well-defined duration may still have an operationalist feel. For, as we shall see below, it means that 
there may be limits to time in cases where scales can be found in the physics, but where no physical process (core of a clock)  can be identified which could in principle exemplify or realize the time scale in question.

Whereas cosmologists often refer to clocks as sketched above, they also
define cosmic time `implicitly' by the specific cosmological model
employed to describe the universe.\footnote{This is related to a more general discussion of the implicit definition of time via natural laws, see Rugh and Zinkernagel (2009), section 2.1.} This can be done e.g. through the relation between time and the scale factor. If we for instance consider a radiation dominated epoch, the Einstein field equations may yield (see e.g. Rugh and Zinkernagel
 (2009), section 4): 
$$ R(t) \propto \sqrt{t} $$
In some sense, the scale factor here serves as a (core of a) clock. However, for this idea to work, one needs to have some bound system or a fixed physical length scale
which does not expand (or which expands differently than the universe). Otherwise, there is no physical content of $R(t)$ and hence no physical content of ``expansion''. Eddington, for
example, emphasized the importance of the expansion of the universe to
be defined relative to some bound systems by turning things upside-down:
`The theory of the ``expanding universe" might also be called the theory
of the ``shrinking atom" (Eddington 1933, quoted from Whitrow 1980, p.
293).

The viewpoint presented here assumes that a physical foundation of
time is closely related to {\em which} physical constituents are available (or at least possible) in the early universe. Such an assumption can be circumvented if  one  subscribes to some sort of Platonism (or mathematical foundationalism) according to which a purely mathematical definition of time, extracted e.g. from the formalism of general relativity (or, as simply the $t$ parameter in some model), is sufficient. According to such a view,  there would seem to be no problem in contemplating, say, periods like $10^{-100}$ seconds  after the big bang.
However, it is widely accepted that the standard cosmological model cannot be extrapolated below Planck scales and, accordingly, that the $t \leftrightarrow \mbox{time}$ interpretation cannot be made for $t$ values below $10^{-43}$
seconds. This illustrates that a physical condition (namely that quantum gravity
effects
may be neglected) can imply a limitation for the $t \leftrightarrow$
time interpretation. But if it is accepted that there is at least {\em
one} physical condition which must be satisfied in order to trust the
backward extrapolation of the FLRW model and its time concept, it
appears reasonable to require that also {\em other} physical conditions
(which are necessary to set up the FLRW model) should be satisfied
during this extrapolation.\footnote{We shall return to this requirement in the final section. We also note that -- except for the
space-time singularity itself -- there are no internal contradictions in
the {\em mathematics} of the FLRW model (or classical general
relativity) which suggests that this model should become invalid at some
point, e.g. at the Planck scale.} Hence, we take it that Platonism is not a satisfactory position regarding time in cosmology. In our view, time has to have some physical basis (i.e. it must be embedded in the available physics) in order to be a well-defined physical concept.

\subsection{The scale problem for time} 

\label{TheScaleProblemForTime}

Let us now shortly review how the above considerations may lead to a scale problem for time in the early universe. The scale problem for time is related to two contemplated phase transitions at $\sim 10^{-5}s$ and $\sim 10^{-11} s$  in the early universe, where the 
notion of length and time scales
(and their physical underpinning in terms of cores of clocks and rods)
 becomes progressively weaker and disappears at $\sim 10^{-11} s$
(if we consider well-known physics set by the standard model of particle physics).

\subsubsection*{$\sim 10^{-5} s$: No bound systems (the quark-hadron phase transition)}

Physically based time and length scales are not independent notions.
Einstein discussed an elementary clock system, `the light-clock', which involves the propagation of a light signal across some physical length scale as when a light signal is being reflected back and forth between the ends of a rigid rod. If we ask whether we in principle may build up such Einstein light-clocks from the constituents of the early universe we note, as we extrapolate backwards in time
(and the temperature rises), that it becomes progressively more difficult to find
any spatially extended physical systems. In the so-called hadron era
there are still bound (hadron) systems such as pions, neutrons and protons. At a transition temperature
of $T \sim 10^{12}$ K ($\sim 10^{-5} s$ after the big bang) it is, however, believed that there
is a quark-hadron phase transition, and above this transition point  no bound states are left. The
universe then consists of particles (like quarks, leptons, gluons and photons) which have no
known spatial extension. If a rudiment (or core) of a rod has to be constructed from a bound
physical system, we no longer have such rudiments of rods left in
the universe, and we have to look elsewhere for physics which can set a
physical length scale.

The quarks and leptons still possess the physical property of
mass. Thus, one still has length scales if the Compton wavelength
$\lambda = \lambda_C = \hbar/(m c)$ of these particles can be taken to
set such a scale. However,  a  rod with
spatial extension equal to the Compton wavelength leads to a
 `pair-production of rods' as a quantum effect (in general, the Compton wavelength is the length scale at which `pair
production' of particle-antiparticle pairs occurs).
It is thus difficult to imagine how the Compton wavelength
divided by $c$ corresponds to a physical process which could function as
the core of a clock e.g. in the above mentioned light-clock.

Note that these considerations, and hence our first proposed time limit at $10^{-5}$ seconds, is based on the somewhat operationalist premise that one should in principle be able to identify a core of a clock (physical process) with a well-defined duration. The contemplated process is one in which a light signal travels a well-defined distance (namely a Compton wavelength of a quark or a lepton), but this process seems physically unrealizable insofar as the photon is converted to a particle-antiparticle pair during flight.\footnote{As different sorts of rudiments (cores) of clocks, we may consider the decay processes of unstable, massive particles such as the decay of the muons $\mu^{-} \rightarrow e^{-} \bar{\nu}_e \nu_{\mu}$, or the decay of the $Z^{0}$ particles, $Z^0 \rightarrow f \bar{f}$ (which can decay into any pair of
fermions). But, as discussed in Rugh and Zinkernagel (2009), also these processes are difficult to conceive as functioning cores of clocks due to their quantum-mechanical and statistical nature.}

\subsubsection*{$\sim 10^{-11} s$: Scale invariance (the electroweak  phase transition)}

According to the standard model of particle physics (which embodies a Higgs sector with
a (set of) scalar field(s) $\phi$) there is an electroweak phase transition at a transition
temperature of $T \sim 300$ GeV $\sim 10^{15}$ K when the universe was
$\sim 10^{-11}$ s old. Above this phase transition the Higgs field
expectation value vanishes $< \phi > = 0$. This transition translates into {\em zero rest
masses} of all the fundamental quarks and leptons (and massive force mediators) in
the standard model. 
Without any masses in the theory it will exhibit a symmetry known as conformal invariance, and it will be impossible to find physical processes (among the microphysical constituents)
with a well-defined (and non-zero) duration.\footnote{See however Rugh and Zinkernagel (2009, section 5.3)  for a brief discussion of some possible rudiments of mass which could remain above the Higgs transition (but which are, in our assessment, insufficient to ground e.g. a physical time scale).}

Thus, not only can no core of a clock be identified. The relevant physics (the electroweak and strong sector)
cannot set physical scales for time, scales for length and no scale for
energy. If there is no scale for length and energy then there is no
scale for temperature $T$. Metaphorically speaking, we may say that not
only the property of mass of the particle constituents `melts away'
above the electroweak phase transition but also the concept of
temperature itself `melts' (i.e. $T$ loses its physical foundation above
this transition point).

In our assessment, therefore, 
the time scale assumed (e.g. in cosmology books) above the electroweak phase
transition is purely speculative in the sense that it cannot be founded
upon an extrapolation of well known physics (due to  conformal
invariance) above the phase transition point. Thus, the time
scale will have to be founded on the introduction of some new physics
(beyond the standard model of particle physics), and is in this sense as
speculative as the new (speculative) physics on which it is based.

It is of interest that Roger Penrose has recently attempted to turn the scale problem for time into a virtue in the construction of a new kind of cosmological scenario. Penrose (2010, p. 142) cites our study\footnote{Our study appeared as a handout in a first print in 2005 (Rugh and Zinkernagel
2005)
and was published in a revised version in 2009.} in connection with the following quote:

\begin{quote}
\noindent
...close to the Big Bang, probably down to around $10^{-12}$ seconds after that moment, when temperatures exceed about $10^{16}$ K, the relevant physics is believed to become blind to the scale factor $\Omega$, and {\em conformal } geometry becomes the space-time structure appropriate to the relevant physical processes. Thus, all this physical activity would, at that stage, have been insensitive to local scale changes. [Emphasis in original]
\end{quote}

\noindent
Our mentioning this point does not imply an endorsement of Penrose's proposal of an ``Extraordinary New View of the Universe" (a conformal cyclic cosmology) in which approximate
conformal invariance holds in both ends (the beginning and the remote
future) of the universe. Nevertheless, there seems to be a certain agreement in philosophical outlook, also when Penrose mentions (2010, p. 93): ``It is important for the {\em physical basis} of general relativity that extremely precise clocks actually exists in Nature, at a fundamental level, since the whole theory depends upon a naturally defined metric {\bf g}" (our emphasis).

\subsubsection*{Why not refer to the Planck scales?}

The combination of the constants $\hbar$ and $c$ from relativistic
quantum mechanics, and $c$ and $G$ from classical general relativity
yields -- as a mathematical combination of physical constants --  the famous Planck scales. As concerns time, the Planck time scale $t_P = (\hbar G/c^5)^{1/2} \sim 10^{-43}$ seconds
is immensely more fine-grained than time scales set by any
physical process which we (in our investigations) have attempted to 
utilize as rudiments of clocks at various stages
in  cosmic history.
{\em If} the Planck time scale  were
 considered sufficient to provide a physical basis for the time scale in the early universe, then there would be no scale problem for time anywhere along the extrapolation from now to the Planck times.

However, we see several related reasons to be suspicious that the Planck time scale does indeed provide a sufficient physical basis for the time scale in the early universe. First of all, the Planck scales are supposed to be the physical relevant scales of theories of quantum gravity, and such theories are still highly speculative. The Planck time scale is therefore at least as speculative as any other imagined time scale above the electroweak phase transition. Second, 
it is expected that quantum
gravity effects are totally negligible at energy scales around the electroweak
phase transition point (and negligible well into the `desert' above this
phase transition). It appears dubious to ground time scales of
Higgs-physics on quantum gravity effects which are irrelevant at Higgs-physics
scales.\footnote{Note
that today we define a second as 9.192.631.770 vibrations
of radiation caused by well-defined transitions in Cs-133 (see  e.g. 't Hooft and Vandoren 2014).
This is a physical grounding (even operationally) of a timescale (a second)
in terms of physical processes taking place at timescales substantially
(10 orders of magnitude) smaller. It would be of interest if 
one could speculate any effect on Higgs-scale
physics stemming from the quantum gravity scales thirty orders of magnitude below.}
Third, even if we bypass the second problem, one may well question how physically reasonable the supposed physical processes would be for grounding the Planck scale (recall that, in our view, a physical basis for a time scale should be related to relevant physical processes). The Planck scales may be  arrived at by setting
the Compton wavelength equal to the Schwarschild radius of a black hole.
It is thus a characteristic scale at which there is a
pair production of black holes as a quantum effect. Consider this in the context of the discussion above on time and length scales in connection with the light-clock: At the Planck length scale there is `pair production of rods' (the rod being the spatial extension of the quantum black hole), and
the corresponding Planck time scale is the time it takes a
light pulse to cross this length scale. This appears
to be more of a mathematical construct than a conceivable physical
process since the crossing of a light pulse is hardly a 
well-defined physical process within such violent fluctuations in the
geometry. For instance, one may ask at which of the two pair produced quantum black
holes the light pulse is supposed to end its `crossing'?\footnote{Note that this third reason against using the Planck scale as a physical basis for the time scale in the early universe is, just like the $10^{-5}$ second limit discussed above, based on the somewhat operationalist premise that we should be able to point to a process (a core of a clock) which provides a definite time interval.}  

In our assessment, then, if  one wants to solve (or dissolve) the scale problem for time at and above the Higgs transition by referring to speculative processes in quantum gravity, such as `quantum pair production of black holes',
then it should at least be admitted that the cosmic time scale constructed in this way
is highly speculative.\footnote{As we shall discuss later,
the {\em  quantum} problem of time (section \ref{TheQuantumProblemForTime})
 does not depend on whether we have a physically well-founded {\em scale} for time. This problem remains (e.g. at the onset of inflation) even if  we base length and time scales (throughout cosmic history) on speculative Planck scale physics.}

\section{Time and world lines}

\label{TimeAndWorldLines}

The above section has focused on the consequences of proper time being related to clocks. As we saw, this relation leads to the idea that time is related to physical processes -- which is a version of what is known as relationism. But there is a more direct route to relationism in cosmology which is independent of the mentioned time-clock-relation (even if in conformity with it). This has to do with the fact that proper time is defined in terms of (possible) particle world lines. In the following we shall discuss how this implies a close relation between time and cosmic matter, both at the global and at the local level.

\subsection{Setting up the FLRW model with a cosmic time}

\label{SettingUpTheFLRWModel}

In Rugh and Zinkernagel (2011) we discuss how the set-up of the FLRW model with a global time is closely linked to the motion, distribution and properties of cosmic matter. We now briefly review some key points of this discussion.

In relativity theory time depends on the choice of reference frame. For the universe, a reference frame cannot be given from the outside, so such a frame has to be ``built up from within", that is, in terms of the (material) constituents within the universe. It is often assumed that the FLRW model 
may be derived just from the cosmological principle. This
principle states that the universe is spatially homogeneous and isotropic (on large scales). It is much less well 
known that another assumption, called Weyl's principle, is necessary in order to arrive at the FLRW model and, in particular, its cosmic time parameter.\footnote{In some cosmology textbooks  -- e.g. by
Bondi, Raychaudhuri and Narlikar -- the importance of Weyl's principle is emphasized, and
explicitly referred to. In other textbooks it appears, in our assessment (see Rugh and Zinkernagel 2011), that the Weyl principle is implicitly assumed in the process of setting up the FLRW model.} Whereas the cosmological principle imposes constraints on the {\em {distribution}} of the matter content of the universe, Weyl's principle imposes constraints on the {\em {motion}} of the matter content. Weyl's principle (from 1923) 
asserts
 that the matter content is so {\em{well behaved}} that a reference frame can be built up from it: 

\begin{quotation}
\noindent
Weyl's principle (in a general form): The world lines of `fundamental
particles' form a spacetime-filling family of non-intersecting geodesics (a congruence of geodesic world lines).
\end{quotation}

The importance of Weyl's principle is that it provides a reference frame which is physically based
on an expanding `substratum' of `fundamental particles' (e.g. galaxies or clusters of galaxies).
In particular, if the (non-crossing)
geodesic world lines are required to be orthogonal to a series of space-like hypersurfaces, a comoving reference frame is defined in which constant spatial coordinates are ``carried by" the fundamental particles (see e.g. figure 3.7 in Narlikar 2002, p. 107).
 The time coordinate is a cosmic time which labels the series of hypersurfaces, and which may be taken as the proper time along any of the particle world lines. We note that the congruence of world lines is essential to the standard cosmological model since the symmetry constraints of homogeneity and isotropy are imposed w.r.t. such a congruence (see e.g. Ellis 1999). Thus, Weyl's principle is {\em{a precondition}} for the cosmological principle; the former can be satisfied without the latter being satisfied but not vice versa.

In the early universe, problems may arise for the Weyl principle and thus for the possibility of identifying a reference frame and a global cosmic time parameter.\footnote{In Rugh and Zinkernagel  (2013) we also argue that there
is no approximate fulfillment of a Weyl principle and no well-defined global (multiverse)
cosmic time concept in the eternal
inflationary multiverse model outlined e.g. by Linde and Guth.}
At present and for most of cosmic history, the comoving frame of reference can be identified as the frame in which the cosmic microwave background radiation (CMB)  looks isotropic 
(see e.g. Peebles 1993, p. 152), and cosmic matter is 
(above the homogeneity scale) assumed to be described as dust particles with zero pressure which fulfill Weyl's principle. But before the release of the CMB, the situation is less straightforward. For, as we go backwards in time, it may become
increasingly more difficult to satisfy, or {\em even formulate}, the Weyl principle
as a physical principle, since the nature of the physical
constituents is changing from galaxies, to relativistic gas particles,
and to entirely massless particles moving with velocity
$c$.\footnote{In the early radiation phase, matter is highly relativistic (moving with  random velocities close to $c$), and the Weyl principle is not satisfied for a typical particle but one may still introduce fictitious averaging volumes in order to create substitutes for `galaxies which are at rest'; 
see e.g. Narlikar (2002, p. 131).}
Indeed, above the electroweak phase transition (before $10^{-11}$ seconds 
`after' the big bang), all constituents are massless and move with 
velocity $c$ in any reference frame. There will thus be no constituents which are comoving  
(at rest).
One might attempt to construct mathematical points (comoving with a reference frame) 
like a center of mass (or, in special relativity, center of energy) 
out of the massless, ultrarelativistic gas
particles, but this procedure seems to require that length scales be available in order to 
e.g. specify how far the particles are apart (which is needed as input in the 
mathematical expression for the center of energy). As discussed earlier, the only option for specifying such length scales (above the
electroweak phase transition) will be to 
appeal to speculative physics, and the prospects of satisfying Weyl's principle
(and have a cosmic time) will therefore 
also rely on speculations beyond current well-established physics. The problem of building up the FLRW model with matter consisting entirely
of consituents moving with velocity $c$, may also be seen by noting that the set-up of 
the FLRW model requires matter (the energy-momentum tensor) to be in the form of 
a perfect fluid, as this is the only form compatible with the FLRW symmetries, 
see e.g. Weinberg (1972, p. 414). For this, a source consisting of pure radiation is not 
sufficient since one cannot effectively simulate a perfect fluid by ``averaging over 
pure radiation".\footnote{Krasinski (1997, p. 5 - 9) notes
that the energy-momentum tensor in cosmological models may contain many different contributions, 
e.g. a perfect fluid, a null-fluid, a scalar field, and an 
electromagnetic field. He also emphasizes that a source of a pure null fluid or a  pure 
electromagnetic field is not compatible with the FLRW geometry, and that solutions with 
such energy-momentum sources have no FLRW limit (Krasinski 1997, p. 13).}

On top of this, the physical basis of the Weyl postulate (e.g.
non-intersecting world lines of `fundamental particles'), and even that of proper time, appears questionable if some period in cosmic history is reached where the
`fundamental particles' are described by wave-functions $\psi (x,t)$
referring to (entangled) quantum constituents. What is a `world line' or
a `particle trajectory' then? (See also the section below on the quantum problem for time). 

In the following we shall briefly question what happens to cosmic time if/when we cannot assume the validity of the standard FLRW model (next subsection), before we turn to the question of what happens if/when the cosmic constituents become quantum (subsection \ref{TheQuantumProblemForTime}).

\subsection{Cosmic time in an inhomogeneous universe }

\label{CosmicTimeInAnInhomogeneousUniverse}

The cosmological standard model is highly idealized and it is therefore of interest to inquire about cosmic time when the model's idealizing assumptions are relaxed. In particular, one may ask whether we still have a good cosmic time concept in our actual -- at least up to very large scales -- inhomogeneous universe? It is well known that close to massive objects time runs differently than in more `void like' segments of spacetime away from any such massive objects. The complexity of constructing a privileged time notion in such situations has been illustrated e.g. in the following example by Feynman. 

How old is our earth? Since clocks (time) run
differently in different gravitational potentials (time dilation in a gravitational field),  time
will run at a different rate
in the center of the earth than on the surface of the earth. Feynman (1995, p. 69) remarks: 
\begin{quotation}
...we might have to be more careful in the future in speaking of the ages of objects
such as the earth, since the center of the earth should be a day or two younger 
than the surface!
\end{quotation}
In fact, the situation is slightly worse,
for integrating up a relative time dilation factor of 
$\bigtriangleup \tau/\tau = \bigtriangleup \Phi/c^2$ over a coarse estimate of the (not precisely defined) lifespan
of our earth ( $\sim 5\times 10^{9}$ years) yields some years in time difference.\footnote{In an
order of magnitude estimate we may assume that our earth is homogeneous and the potential
difference between the center and the surface of our earth is then integrated up to
$\bigtriangleup \Phi = G M/2 R \;$ which translates into a relative time dilation effect 
$\bigtriangleup \tau/\tau = \bigtriangleup \Phi/c^2 = 1/4 \times (R_{Schw}/R) \sim 1/3\times
10^{-9}$ (Here $R_{Schw} = 2GM/c^2$ is the Schwarzschild radius of our earth with
mass $ M $ and radius $ R $). Integrating this relative time dilation over $\sim 5 \times 10^9$ years
yields an order of magnitude estimate of $\sim 2$ years for the age difference (as measured by
counterfactual clocks located in the center and at the surface of our earth
 over the lifespan of our earth).}

In a universe with an inhomogeneous distribution of
the material constituents, the situation is less clear than in
the Feynman example of a slightly inhomogeneous gravitational field 
throughout our earth. In some mathematically simplified spatially inhomogeneous models, it
may be possible to maintain a Weyl principle and a notion of a global cosmic time
(cf. e.g. Krasinski (1997) and references therein).
However, if our universe exhibited fractal behavior and collisions on all scales 
it would  be difficult to uphold a Weyl principle (even in an
`average sense' where small scale 
collisions and inhomogeneities are averaged out). We may add that such fractal behavior and collisions on all
scales appear to be a characteristic of envisaged multiverse inflationary scenarios like
chaotic inflation, see e.g. discussion and
references in Rugh and Zinkernagel (2013).

Not least due to the observed microwave background isotropy
(and the remarkable isotropy of X-ray counts, radio source counts, and $\gamma$-ray bursts)
it is generally expected (yet debated)
among cosmologists that there will be a transition from small-scale fractal behavior to
large-scale homogeneity.\footnote{Moreover, it has been emphasized, e.g. by Barrow (2005), 
that large contrasts in density $\delta \rho/\rho$
are not necessarily  mirrored in similar inhomogeneities in the
gravitational potential $\Phi$ since the equation of the
relative perturbation $\delta \Phi /\Phi$
of the gravitational potential has in it  a huge suppression factor 
($\delta \Phi/\Phi \sim \delta \rho/\rho \times (L/(c/H))^2$) if the size $L$ of the
 density irregularity is small relative to the Hubble radius $c/H$.}
A recent study arguing this case is e.g.
 Scrimgeour et al. (2012).\footnote{If we want to observationally test the expected homogeneity at large
scales, one should pay attention to the danger of vicious circularities (``catch-22").
Distance measures like redshift-distance measures (at large distances)
should not have built-in the assumptions we want to test (the FLRW model as space-time metric, etc.).
The analysis provided in e.g.  Scrimgeour et al. (2012) is very elaborate but it
is of interest that they
note (p. 4): ``To do this, we assume the FRW metric and
$\Lambda$CDM. This is necessary for any homogeneity measurement, since
we must always assume a metric in order to interpret redshifts. Therefore in the
strictest sense this can {\em only} be used as a consistency test of 
$\Lambda$CDM. However, if we find the trend towards homogeneity matches the
trend predicted by $\Lambda$CDM, then this is a strong consistency check for
the model and one that an inhomogeneous distribution would find difficult to
mimic" (our emphasis).}
Nevertheless, even if our universe is not fractal at the largest -- but only at intermediate -- distance
scales, it is an interesting question how significantly this may change the cosmic time concept of the resulting cosmological model. Indeed, inhomogeneous models with a fractal matter
distribution at intermediate scales will presumably
exhibit more complicated conceptions of
cosmic time than in the highly symmetric, idealized FLRW model universes.

One way to address what happens in an inhomogeneous universe,
is to attempt to construct a notion of cosmic time associated with an event
(here, now) by looking at the proper time (e.g. MTW (1973), \S 13.4 \S 27.4)
\begin{equation} \label{propertimeworldline1}
\tau = \tau (\gamma ) = \int_{\gamma} \sqrt{g_{\mu \nu} dx^{\mu} dx^{\nu}} \; \;,
\end{equation}
along (particle) timelike world lines (indicated with subscript $\gamma$ and with
4-velocities $u^{\mu} (v) = dx^{\mu} (v)/d v$),  which starts at the beginning of space time, and ends in the
event (here, now).\footnote{Such a definition is only well-defined, i.e. the proper time is only finite, if there is a beginning of space-time e.g. in a ``big bang" (see also Lachi\`eze-Rey 2014, section 5.3), or if we chose some arbitrary starting point (assumed to exist) from which we can integrate (Ellis 2012, section 3).}  
But along which world lines $\gamma$ should the proper time integral be taken?

Ellis (2012, p. 9-10)  proposes to take the proper time integral along a specific set of preferred fundamental world lines, which (for realistic matter) are uniquely geometrically
determined.
This construction does not invalidate the Weyl principle but rather builds on it and develops it (Ellis, private communication).\footnote{We note that Ellis assumes that there is a uniquely defined
vector field of 4-velocities $u^{\mu} (v) = d x^{\mu}(v)/dv $  (if such 4-velocities
are uniquely defined in each spacetime point on the manifold, this is equivalent
to assuming the existence of a congruence of world lines which
are non-crossing). According to Ellis' proposal, these 4-velocities 
(in order to be preferred fundamental world lines) should
satisfy that they are timelike eigenlines of the Ricci tensor,
$R_{\mu \nu} u^{\nu} = \lambda u_{\mu}$.} The `present' is in this
construction defined as the surface $\left\{ \tau = constant \right\}$ determined by
taking the proper time integral (\ref{propertimeworldline1})
over the family of fundamental world lines starting at the
``big bang". However, according to Ellis, the equal time hypersurfaces can in generic situations be much more complicated (see discussion in Ellis 2012, p. 10) than the simple equal (cosmic) time hypersurfaces in the FLRW model universes. In particular, Ellis remarks that the equal time hypersurfaces may not even necessarily be spacelike in an inhomogeneous spacetime. 

It therefore appears to
be a complicated -- and to our knowledge still open -- question whether the resulting concept of cosmic time exhibits the properties which allow for a `backward' extrapolation into an `early' inhomogeneous universe.

\subsection{The quantum problem for time}

\label{TheQuantumProblemForTime}

We have seen above that it may be difficult to identify a {\em{global}} cosmic time (without the Weyl principle), and in earlier sections also that there may not be a {\em {scale}} for time (before the Higgs transition). Even if this is so, it might still be possible to maintain a local time {\em order}, i.e to ask about the past of some particular event -- for instance, the past of the onset of inflation. However, as we shall indicate below, it may well be that not even a {\em{local}} (and scale-free) time order is available as time is extrapolated backwards in the very early universe.

The origin of this local (or quantum) problem for time is due to the widely assumed ``quantum fundamentalist" view  according to which the material constituents of the universe could be described {\em exclusively} in terms of quantum theory at some early stage of the universe. Such a perspective is natural in quantum cosmology (and quantum gravity), in which spacetime itself is treated 
quantum mechanically (see also Hartle 1991). From the point of view of such theories, it has been argued  that a quantum problem of time appears already (in the backward extrapolation from now) at the onset of inflation. Thus, Kiefer (e.g. 2003) affirms that:

\begin{quote}
\rightskip=0pt
The Universe was essentially ``quantum" at the onset of inflation. Mainly due to bosonic fields, decoherence set in and led to the emergence of many ``quasi-classical branches" which are dynamically independent of each other. Strictly speaking, the very concept of time makes only sense after decoherence has occurred. In addition to the horizon problem etc., inflation also solves the “classicality problem”. [...]
Looking back from our Universe (our semiclassical branch) to the past, one would notice that at the time of the onset of inflation our component would interfere with other components to form a timeless quantum-gravitational state. The Universe would thus cease to be transparent to earlier times (because there was no time).
\end{quote}

The problem here seems to be that our spacetime (and therefore time) `dissolves' into a superposition of spacetimes at the onset of inflation, and in this sense Kiefer acknowledges a quantum problem of time at this point. The situation, however, might be worse (i.e. the quantum problem may appear earlier in a backward extrapolation from now), since the appeal to decoherence is questionable. To see this, consider what one might call the cosmic measurement problem, which addresses the quantum mechanical measurement problem in a cosmological context:

\begin{quotation}
\noindent
{\em The cosmic measurement problem}: 
If the universe, either its content or in its entirety, was once (and still is) 
quantum, how can there be (apparently) classical structures now? 
\end{quotation}

While many aspects of the cosmic measurement problem have been addressed in the literature, the perspective which we have tried to add is that the problem is closely related to providing a physical  basis for the (classical) FLRW model with a (classical) cosmic time parameter. As illustrated in the Kiefer quote above, an often attempted response to the cosmic measurement problem is to proceed via the idea of decoherence. According to this idea, some degrees of freedom are regarded as irrelevant (they are deemed inaccessible to measurements and are traced out), and they are therefore taken to act as an environment for the relevant variables. The picture is that the environment in a sense `observes' the system in a continuous measurement process and thus suppresses superpositions of the system (see e.g. Kiefer 1989).

However, as is widely known, decoherence cannot by itself solve the measurement problem and explain the emergence of a classical world.\footnote{For a simple explanation of this, and some references to
the relevant literature, see e.g. Zinkernagel (2011, section 2.1).} For, if both environment and system are quantum, the total state of the system (relevant plus irrelevant degrees of freedom) is still a superposition. According to quantum mechanics, no definite (classical) state can therefore be attributed to any of the components. As argued by Sudarsky 
(2011, section 4.1), this problem is only aggravated in the cosmological context since one cannot here appeal to the usual pragmatic considerations regarding what classical observers and their measurement apparatus would register.\footnote{From a pragmatic point of view, quantum mechanics may be seen as a theory of expected outcomes of measurements, in which both apparatus and observers are kept outside the quantum description. We have pointed out elsewhere (Rugh and Zinkernagel 2005, Zinkernagel 2016) that Bohr went beyond this pragmatic (or instrumental) interpretation. His view was rather a contextual one according to which any system can be treated quantum mechanically but not all systems can be treated this way at the same time.} In spite of such worries, Kiefer (2003) contemplates that decoherence successively classicalizes different constituents of the universe: At the onset of inflation, the inflaton field itself is classicalized and, at the end of inflation, decoherence converts the quantum fluctuations of the inflaton field into classical density perturbations (seeds of structure).\footnote{In this regard, Anastopoulos (2002) mentions a worry about decoherence closely related to the ones already noted: ``\dots a sufficiently classical behavior for the environment seems to be necessary if it is to act as a decohering agent and we can ask what has brought the environment into such a state ad-infinitum".} 

But even if one were to bypass the strong arguments against decoherence as a solution to the cosmic measurement problem, a potentially more serious problem is lurking: If decoherence is to explain the emergence of classical structures, it cannot -- as in environmentally induced decoherence -- be a process in (cosmic) time,
insofar as classical structures (particle world lines) are needed from
the start to define time both locally and globally! There thus seems to be a {\em vicious circularity} if one invokes decoherence to explain the `emergence' of time, which we can formulate in slogan form:

\begin{quote}
{\em{Decoherence takes time and cannot therefore provide time.}}
\end{quote}

\noindent This implies that several of the temporal expressions in the quote by Kiefer given above (``decoherence {{\em sets in}}", ``{\em after} decoherence {\em has occurred}", etc.) are strictly speaking without meaning. 

Although the discussion above has focused on decoherence, we note that the quantum problem of time seems to be shared by other ``quantum fundamentalist" views even when these do not rely essentially on decoherence (e.g. the spontaneous collapse model described in Sudarsky 2011). Our point is that any interpretation of quantum mechanics will need a time concept -- which is bound up with
the notion of possible (classical) particle world lines -- in order to address the early universe. The assumption of a quantum nature of the material (or otherwise) constituents of the universe makes it hard  (or impossible) to associate these with well-defined particle trajectories. During inflation the only relevant constituent of the universe is taken to be the inflaton field $\varphi$ which -- in the last analysis -- is a quantum field.  And just as wave functions in non-relativistic quantum theory do not give rise to physical motion (of a particle or wave) in space and time -- without assumptions solving the measurement problem -- so quantum fields do not describe moving elementary particles in space with well-defined trajectories.

Up to this point we have discussed the quantum problem of time from a quantum fundamentalist point of view based on quantum cosmology or quantum gravity. Let us now proceed from the present (and more cautious) perspective, in which we start from a classical point of view and attempt to extrapolate proper time backwards. More specifically, consider the past of some event by extrapolating backwards the proper time integral along a world line 
with
4-velocity $u^{\mu}(v) = dx^{\mu} (v)/d v$
which ends in the event (formula as in
equation (\ref{propertimeworldline1}) in section 
\ref{CosmicTimeInAnInhomogeneousUniverse}).
This approach assumes that we know the metric and that there are well-defined 4-velocities. The question then becomes whether such 4-velocities (or, equivalently, world lines) can always be constructed, i.e. physically realized as opposed to merely mathematically defined, from the available constituents (e.g. from a scalar field $\varphi$)?\footnote{One idea here would be to equate the energy momentum tensor of the perfect fluid form with the energy momentum tensor for the scalar field. This results in the 4-velocity 
$u_{\mu} = A \cdot \partial_{\mu} \varphi$ where $A = (\partial^{\nu} \varphi \; \partial_{\nu} \varphi)^{-1/2}$, see e.g. Krasinski (1997, p. 8) and Hobson et al. (2006, p. 432). }

In the inflationary scenario, the relevant candidate for constructing sensible notions of particle world lines and classical trajectories will have to come from the $\varphi$ field. And even if we were allowed to take this field as effectively classical (described by the lowest order approximation in quantum field theory) during inflation (e.g. during a slow-roll evolution), the quantum problem of time will be faced at the on-set of inflation. At this point (supposed to be the `birth' of our bubble universe in a multiverse setting), the inflaton field is strongly quantum: Quantum fluctuations with amplitudes (within a factor of ten) of the order of the Planck scale are necessary to reset or lift the scalar field to a value where a new bubble (our universe) is born and becomes dominated by inflation (see e.g. Linde 2004, section 4).
Thus, at the beginning of inflation (or the `birth' of our universe), the $\varphi$ field is nowhere close to being a classical field on top of which we have small quantum fluctuations. Rather, it is entirely dominated by Planck scale quantum fluctuations.

In summary, to use the local time concept for contemplating times before inflation (or, indeed, earlier bubble-universes in the multiverse), it must be possible to identify (or, at least, to speculate) a particle world line along which proper time can be extrapolated backwards.\footnote{From our relationist point of view -- in which time is necessarily related to physical processes -- the time-like curves can only be identified (they only have a physical basis) if the motion of objects or test particles along these curves is at least in principle realizable given the available physics.} But, as we have seen, it is unclear how one would go about constructing any individual classical particle world line from the inflationary scalar field $\varphi$ in a regime where its quantum behaviour is dominant (at the onset of inflation).  If such world lines (classical trajectories) cannot be constructed from the underlying physics (the $\varphi$ field), it seems that the very conditions for speaking about the past of an event in general relativity are not fulfilled. Hence, in our assessment, a pure quantum phase in the early universe implies that proper time (and even its order aspect, that is, its ability to distinguish before and after) is no longer a well-defined concept.

\section{Summary and discussion}

\label{SummaryAndDiscussion}

It is common practice to extrapolate the standard cosmological model back to at least the Planck time. In this manuscript, we have tried to insist that this is problematic. The underlying philosophical reason is that the extrapolation of
the FLRW model and its time concept requires, in our view, that
 the physical basis of time in the model and, more
generally, the physical conditions needed to set up the model, 
are not invalidated along this extrapolation. 
This situation gives rise to a number of possible limits of time, respectively, at $\sim 10^{-5} s$, $10^{-11} s$, $10^{-34} s$, and $10^{-43} s$ `after' the mathematical point $t = 0$ in the FLRW model. 

As briefly hinted in section \ref{TimeAndClocks}, we are aware that we are here making a philosophical choice -- at least concerning the two first limits. For we are assuming that the natural laws need a physical basis at all points along the extrapolation, as opposed to just having a basis at the present epoch (when it is easy to identify not only length and time scales, but also physical processes with well-defined durations). The difference between the first two limits and the Planck time (and possibly the time of onset of inflation) is that the former two (phase transitions) do not mark events where the natural laws are expected to break down. Rather, the two phase transitions are predictions of the natural laws themselves (by contrast, classical gravity is expected to break down at the Planck scale). Hence, in the case of the first two time limits, the problem concerns the {\em interpretation} of the natural laws; i.e. whether we are entitled to interpret the laws as physical laws throughout the backward extrapolation, if the foundation for this interpretation (like the existence of cores of rods and clocks) disappears at some point along the extrapolation.
Given our view of the interpretation of natural laws, the time concept in the early universe becomes speculative before the electroweak phase transition. As we have seen, before this point ($\sim 10^{-11}$ seconds), known physics becomes scale invariant and so one loses any (non-speculative) handle on how close we are to the singularity. We believe, but it should be further examined, that our position is a reasonable compromise between Platonism (mathematical foundationalism) and operationalism (which requires a method for actually measuring cosmic time).

In sections \ref{SettingUpTheFLRWModel} and 
\ref{CosmicTimeInAnInhomogeneousUniverse}
 we have seen that a {\em{global}} concept of cosmic time (with or without a scale) may become problematic in the early universe if the Weyl principle cannot be satisfied (e.g. if everything moves with the speed of light and no comoving reference frame can be constructed). Moreover, the discussion in section 
\ref{TheQuantumProblemForTime}
 showed that not even a {\em {local}} concept of time, which could be used to address the past of some local event, may be available as time is extrapolated backwards in the very early universe. In particular, this seems to be the case if one assumes ``quantum fundamentalism"; the idea that everything is quantum, and even if something looks classical {\em now}, there was an early time, e.g. $10^{-34}$ seconds, when nothing did. Thus, if all constituents are quantum at the onset of inflation $10^{-34}$ s, it seems difficult (or impossible) to even construct a physical notion of proper (local) time along individual world lines which could order events in the very early universe.
The upshot of our discussion on these points was that classical systems 
appear to be necessary throughout cosmic history (to have a reasonable time concept). It is standard to hold that  quantum gravity sets in at $10^{-43}$ s,
i.e. that there is no time concept ``before" this Planck time. But our discussion indicates that {\em{if}} one believes that everything is quantum, then one has a problem with time in general (and not only in quantum gravity)!

Let us finally briefly consider whether the possible limits to time are a misfortune for cosmology. We think not. Limits in science are good for at least two reasons. First, they should not be seen as stumbling blocks for research but rather as invitations to keep asking questions, e.g. as to which theories might describe what lies beyond the present temporal limits (or how the limits
might be circumvented, e.g. by introducing speculative new physics).
Such invitations can be expected to remain open 
since for any postulated theory describing earlier times, it will probably always be possible to ask: what  lies beyond {\em that} theory? 
This leads to the second reason: The fact, if it is a fact, that there will always be something beyond our (current?) scientific understanding may be aesthetically attractive, if not also comforting.\footnote{See Zinkernagel (2014) for some brief remarks on aesthetics in cosmology.} Both of these reasons for endorsing limits are connected to that feeling of wonder which has been an important driving force throughout the history of cosmology.

\section*{Acknowledgements}

We are grateful for discussions on the above topics with many people
over the years (cf. acknowledgements in previous manuscripts). We also
thank the organizers of the Philosophy of Cosmology conference at
Tenerife for the opportunity to present this work. HZ thanks the Spanish
Ministry of Science and Innovation (Project FFI2011-29834-
C03-02) for financial support.

\section*{References}

\noindent
-- Anastopoulos, C. 2002. Frequently asked questions about 
decoherence. {\em International Journal of Theoretical Physics}, {\bf 41}, 2002, 1573--1590. 

\noindent
-- Barrow, J.D. 2005. Worlds without end or beginning. In D. Gough (ed)
{\em The Scientific Legacy of Fred Hoyle}, 93--101. Cambridge: Cambridge University Press.

\noindent
-- Bondi, H. 1960. {\em Cosmology} (2nd edition). Cambridge: Cambridge University Press.

\noindent
-- Ellis, G.F.R. 1999. 83 years of general relativity and cosmology: progress and problems. {\em Classical and Quantum Gravity}, {\bf 16}, A37--A75.

\noindent
-- Ellis, G.F.R.  2012.  Space time and the passage of time,
gr-qc arXiV: 1208.2611.

\noindent
-- Feynman, R.P. 1995. {\em  Lectures on Gravitation} (edited by B. Hatfield). Addison-Wesley.

\noindent
-- Hartle, J.B. 1991. The Quantum Mechanics of Cosmology. In S. Coleman et al (eds.)
{\em Quantum Cosmology and Baby Universes}. Singapore: World Scientific.

\noindent
-- Hobson, M. P., Efstathiou, G. P. and Lasenby, A. N. 2006.
{\em General Relativity}. Cambridge University Press.

\noindent
-- Kiefer, C. 1989. Continuous measurement of intrinsic time by fermions.
{\em Classical and Quantum Gravity}, {\bf 6}, 561--566.

\noindent
-- Kiefer, C. 2003. Decoherence in Quantum Field Theory and Quantum Gravity.
In E. Joos et al (eds) {\em Decoherence and the Appearance of a Classical World in
Quantum Theory}, 181--225. Berlin: Springer.

\noindent
-- Krasinski, A. 1997. {\em Inhomogeneous Cosmological Models}. Cambridge: Cambridge University Press.

\noindent
-- Lacheize-Rey, M.  2014. In search of relativistic time.
 {\em  Studies in History and Philosophy of Modern Physics}, {\bf 46}, 38--47.

\noindent
-- Linde, A. 2004.  Inflation, quantum cosmology, and the anthropic principle. In J. D. Barrow, P. C. W. Davies and C. L. Harper, (eds) {\em Science and
Ultimate Reality},  426--458. Cambridge: Cambridge University Press. 

\noindent
-- Misner, C. W., Thorne, K., and Wheeler, J. A. 1973. {\em
Gravitation}. New York:  W. H. Freeman.

\noindent
-- Narlikar, J. V. 2002. {\em An Introduction to Cosmology}, (third
edition). Cambridge: Cambridge University Press.

\noindent
-- Peacock, J. A. 1999.  {\em Cosmological Physics}. Cambridge: Cambridge
University Press.

\noindent
-- Penrose, R. 2010.  {\em Cycles of Time - An Extraordinary New view of the Universe}. London: Random House. 

\noindent
-- Peebles, P. J. 1993.  {\em Principles of physical cosmology}. Princeton: 
Princeton University Press.

\noindent
-- Raychaudhuri, A. K. 1979.  {\em Theoretical Cosmology}. Oxford: Clarendon Press.

\noindent
-- Rugh, S. E. and Zinkernagel, H.  2005.  Cosmology and the Meaning of Time, 76pp. 
Distributed manuscript. 

\noindent
-- Rugh, S. E. and Zinkernagel, H. 2009.  On the physical basis of cosmic time. 
{\em Studies in History and Philosophy of Modern Physics}, {\bf 40}, 1--19. 

\noindent
-- Rugh, S. E. and Zinkernagel, H.  2011.  Weyl's principle, Cosmic Time and Quantum
Fundamentalism. In D. Dieks et al (eds) {\em Explanation, Prediction and Confirmation. The Philosophy of Science in a European Perspective}, 411--424. Berlin: Springer.

\noindent
-- Rugh, S. E.  and Zinkernagel, H. 2013. A Critical Note on Time in the Multiverse, 
in V. Karakostas and D. Dieks (eds) {\em Recent Progress in Philosophy of Science: Perspectives and Foundational Problems}, 267--279. Berlin: Springer. 

\noindent
-- Scrimgeour, M. I. et al.  2012.  The WiggleZ Dark Energy Survey: The 
transition to large-scale cosmic homogeneity, arXiv: 1205.6812v2.

\noindent
-- Sudarsky, D. 2011.  Shortcomings in the Understanding of Why Cosmological Perturbations Look Classical. {\em International Journal of Modern Physics D}, {\bf  20}, 509--552.

\noindent
-- 't Hooft, G. and Vandoren, S. 2014.  {\em Time in Powers of Ten}.
World Scientific.

\noindent
-- Weinberg, S. 1972. {\em Gravitation and Cosmology.} New York: Wiley and
Sons.

\noindent
-- Whitrow, G. J. 1980.  {\em The natural philosophy of time.} Oxford:
Clarendon Press.

\noindent
-- Zinkernagel, H. 2008.  Did Time have a Beginning? {\em International
Studies in the Philosophy of Science}, {\bf 22} (3), 237--258.

\noindent
-- Zinkernagel, H.  2011.  Some trends in the philosophy of physics. 
{\em Theoria}, {\bf 26} (2), 215--241.

\noindent
-- Zinkernagel, H. 2014.  Introduction: Philosophical aspects of modern cosmology.
{\em  Studies in History and Philosophy of
Modern Physics}, {\bf 46}, 1--4.

\noindent
-- Zinkernagel, H. 2016.  Niels Bohr on the wave function and the
classical/quantum divide. {\em Studies in History and Philosophy of Modern Physics}, {\bf 53}, 9--19.

\end{document}